\documentclass[aps,prl,reprint,twocolumn,superscriptaddress]{revtex4-1}
\usepackage{amsmath}
\usepackage{amssymb}
\usepackage{natbib}

\newcommand {\e} {\varepsilon}

\def \w {\omega}
\def \W {\Omega}
\def \vp {\varphi}

\def \ii {\text{i}}

\newcommand{\avr}[1]{\langle #1 \rangle}

\newcommand{\be}{\begin{equation}}
\newcommand{\ee}{\end{equation}}

\newcommand{\bea}{\begin{eqnarray}}
\newcommand{\eea}{\end{eqnarray}}

\begin{document}
\title{Dynamics of weakly inhomogeneous oscillator populations: Perturbation
theory on top of Watanabe-Strogatz integrability}

\author{Vladimir Vlasov}
\affiliation{Center for Neuroscience and Cognitive Systems,
Istituto Italiano di Tecnologia, Corso Bettini, 31, I-38068 Rovereto, Italy}
\affiliation{Institute for Physics and Astronomy, University of Potsdam,
14476 Potsdam, Germany}
\author{Michael Rosenblum}
\affiliation{Institute for Physics and Astronomy, University of Potsdam,
14476 Potsdam, Germany}
\author{Arkady Pikovsky}
\affiliation{Institute for Physics and Astronomy, University of Potsdam,
14476 Potsdam, Germany}
\affiliation{Department of Control Theory, Nizhni Novgorod State University,
  Gagarin Av. 23, 606950, Nizhni Novgorod, Russia}

\date{\today}
\begin{abstract}
As has been shown by Watanabe and Strogatz (WS) [Phys. Rev. Lett., 70,
2391 (1993)], a population of identical phase oscillators, sine-coupled to a common field,
is a partially integrable system: for any ensemble size its dynamics reduces to 
equations for three
collective variables. Here we develop a perturbation approach for weakly nonidentical ensembles.
We calculate corrections to the WS dynamics for two types of perturbations: 
due to a distribution of natural frequencies and of forcing
terms, and due to small white noise. We demonstrate,  that in both cases the
complex mean field for which the  dynamical equations are written, is close to 
the Kuramoto order parameter, up to the leading order
in the perturbation. This supports validity 
of the dynamical reduction suggested by Ott and Antonsen [Chaos, 18, 037113 (2008)]
for weakly inhomogeneous populations.
\end{abstract}

\maketitle
	
%-------------------------------------------------
Dynamics of oscillator populations arouses large interest across 
different fields of science 
and engineering~\cite{Strogatz-00,Acebron-etal-05,Pikovsky-Rosenblum-15}. 
Relevant physical examples are arrays of Josephson junctions or
lasers, metronomes on a common support, 
ensembles of electronic circuits, spin-torque, 
 optomechanical and electrochemical
oscillators~\cite{Wiesenfeld-Colet-Strogatz-96,*Nixon_etal-13,%
*Martens_etal-13,*Temirbayev_etal-12,*Grollier-Cros-Fert-06,%
*Heinrich_etal-11,*Kiss-Zhai-Hudson-02a},  
etc.
The concept of coupled oscillator
populations finds also broad application in life
sciences, in particular in neuroscience
~\cite{Richard-Bakker-Teusink-Van-Dam-Westerhoff-96,%
*Breakspear-Heitmann-Daffertshofer-10,*Prindle_etal-12},
and even in description of social 
phenomena~\cite{Neda-Ravasz-Brechet-Vicsek-Barabasi-00,Strogatz_et_al-05}. 
The paradigmatic model in this field is the Kuramoto model of globally 
coupled phase oscillators~\cite{Kuramoto-75,*Kuramoto-84,Sakaguchi-Kuramoto-86}.
Remarkably, this setup and its simple generalizations explain not only the 
emergence of collective mode, which can be viewed as a
nonequilibrium disorder-to-order
transition~\cite{Gupta-Campa-Ruffo-14}, 
but also many other interesting dynamical phenomena
like partial synchrony~\cite{Rosenblum-Pikovsky-07,*Pikovsky-Rosenblum-09} and
chimera states~\cite{Kuramoto-Battogtokh-02,Abrams-Mirollo-Strogatz-Wiley-08}.

A striking property of a system of  $N$ identical oscillators,
sine-coupled to a common field  (with the famous Kuramoto and 
the Kuramoto-Sakaguchi
models being representatives of this class),  is the 
partial integrability of the system
for $N>3$, established in the seminal work by Watanabe and
Strogatz (WS)~\cite{Watanabe-Strogatz-93,*Watanabe-Strogatz-94}.
The WS theory (below we explain it in sufficient details) allows one to reduce 
the dynamics of $N$ oscillators to 
that of three collective variables 
and $N-3$ constants~\cite{Pikovsky-Rosenblum-08,Marvel-Mirollo-Strogatz-09,
Pikovsky-Rosenblum-11}, see also recent review~\cite{Pikovsky-Rosenblum-15},
and is valid for arbitrary common force, which can be, 
e.g., stochastic~\cite{Braun-etal-12}. 
The identity of oscillators is essential, as well as the restriction that 
all units are forced equally.
%(Thus, e.g., common noise does not violate the integrability conditions). 
However, natural systems always possess at least a small degree of
inhomogeneity, and the goal of this Letter is to extend
the WS approach to cover this case.

To achieve this goal 
%by developing a perturbation theory on top of the WS integrability. 
we first re-write the original equations of the inhomogeneous oscillator 
population in a equivalent, but suitable for a perturbation analysis, form. 
(This can be considered as an analogy to writing Hamiltonian equations 
in action-angle variables and thus bringing them to a form, ready  
for an approximate analysis.) 
Next, in the limit of small inhomogeneity we approximately reduce the 
dynamics to weakly perturbed WS equations with a certain 
distribution of WS constants. 
For an illustration of our theory we analyze two particular problems: 
(i) an ensemble with distributions of the natural frequencies and of the forcing, 
i.e. the case of a quenched (time-idependent) disorder, and (ii) stochastic
perturbations due to uncorrelated white-noise terms acting on 
all units.
For both examples we derive in the thermodynamic limit $N\to\infty$
corrections to the WS equations as well as the relation between 
the Kuramoto order parameter and the WS complex amplitude in the leading order 
in the perturbation amplitude. 

We start with general equations for $N$ sine-coupled phase oscillators 
\be
\label{eq:posc}
\dot\vp_k=\w(t)+{\rm Im}\left[H(t) e^{-\ii\vp_k}\right]+
F_k \;,\quad k=1,\ldots,N\;.
\ee
Here $\w$ and $H$ describe general time-dependent 
common forcing. In particular, $H$ can depend on the mean field,
like in the Kuramoto setup (see~\cite{Vlasov-Pikovsky-Macau-15} for an example 
where mean field enters $\w$).  
$F_k$ are general inhomogeneous terms that also can be 
time- and $\vp_k$-dependent.
Notice that the case $F_k=0$ is solved by the WS theory.
 
We re-write Eq.~(\ref{eq:posc}) as 
\begin{equation}
\frac{d}{dt}\left(e^{\ii\vp_k}\right)=\ii[\w(t)+F_k]e^{\ii\vp_k}
+\frac{1}{2}H(t)-\frac{e^{\ii 2\vp_k}}{2}H^*(t)
\label{eq:expform}
\end{equation}
and perform the M\"obius transformation~\cite{Marvel-Mirollo-Strogatz-09} from $\vp_k$
to the WS complex amplitude $z$ and new WS phases $\psi_k$, according to
\be
\label{eq:mtr}
e^{\ii\varphi_k}=\frac{z+e^{\ii \psi_k}}{1+z^*e^{\ii \psi_k}}\;.
\ee
Next, we search for the solution for $z$ in form of the WS equation
(cf. Eq.~(10) in \cite{Pikovsky-Rosenblum-08})
with an additional complex perturbation term $P$, to be determined later: 
\be
\label{eq:newz}
\dot{z}=\ii\omega z+ \frac{H}{2}-\frac{H^*}{2}z^2 + P \;. 
\ee
Substituting Eqs.~(\ref{eq:posc},\ref{eq:mtr},\ref{eq:newz}) into Eq.~(\ref{eq:expform})
we obtain for the WS phases
\be
\label{eq:phased}
\begin{aligned}
\dot{\psi}_k=\omega +&{\rm Im}(z^* H)+
 F_k\left[ \frac{2\text{Re}
\left(ze^{-\ii\psi_k}\right)+1+|z|^2}{1-|z|^2}\right]\\
&-\frac{2{\rm Im}\left[P\left(z^*+e^{-\ii\psi_k}\right)\right]}{1-|z|^2}\;.
\end{aligned}
\ee

Since the M\"obius transformation \eqref{eq:mtr} from 
$\vp_k$ to the set $(\psi_k,z)$
is under-determined, we impose one complex 
condition to ensure uniqueness of determination of the
complex variable $z$.
Following Refs.~\cite{Watanabe-Strogatz-93,*Watanabe-Strogatz-94,Pikovsky-Rosenblum-08,%
Marvel-Mirollo-Strogatz-09,Pikovsky-Rosenblum-11} 
we require: 
\be\label{eq:fhzero}
\frac{1}{N}\sum_{k=1}^Ne^{\ii\psi_k}=0\;.
\ee
If this condition is valid at $t=0$ for the initial
M\"obius transformation, then it will be valid at all times
provided $\frac{d}{dt}\sum_k \exp[\ii \psi_k]=0$. 
Substituting here Eq.~\eqref{eq:phased}, we obtain
\be
\label{eq:b1}
P-P^*U=
\frac{\ii}{N}\sum_{k=1}^N
F_k\left[z+(1+|z|^2)e^{\ii\psi_k}+z^*e^{2\ii\psi_k}\right],
\ee
where
$U=N^{-1}\sum_{k}\exp[{\ii 2\psi_k}]$.
Together with the complex conjugate of Eq.~\eqref{eq:b1}, this 
allows us to express $P$:
\be\label{eq:b}
\begin{aligned}
P=&
\frac{\ii}{(1-|U|^2)N}\sum_{k=1}^N
F_k\left[z(1-Ue^{-2\ii\psi_k})+\right.\\
+&\left. (1+|z|^2)(e^{\ii\psi_k}-
Ue^{-\ii\psi_k})+z^*(e^{2\ii\psi_k}-U)\right]\;.
\end{aligned}
\ee
Equations~(\ref{eq:newz},\ref{eq:phased},\ref{eq:b})
constitute a closed system for new variables $z,\psi_k$.
We emphasize that our derivation of Eqs.~(\ref{eq:newz},\ref{eq:phased},\ref{eq:b})
is exact; the only restriction is that $|z|\neq 1$ and $|U|\neq 1$,
i.e. the cases of full synchrony and of two perfect clusters are 
excluded.

As one can see from Eq.~\eqref{eq:b}, for the
non-perturbed case $F_k=0$ the perturbation term $P$  vanishes 
and one obtains the WS 
equations~\cite{Watanabe-Strogatz-93,*Watanabe-Strogatz-94,Pikovsky-Rosenblum-08}.
In the standard formulation, for $F_k=0$ we have $\dot{\psi}_k=\omega +{\rm Im}(z^* H)$
and therefore one can introduce variable $\alpha$, satisfying 
$\dot\alpha=\omega +{\rm Im}(z^* H)$, and constants 
$\bar\psi_k=\psi_k-\alpha$, what completes the WS equations, see
\cite{Pikovsky-Rosenblum-08}. 

It is instructive to discuss physical meaning of the WS
variables. The WS complex amplitude
$z=\rho e^{\ii \Phi}$, $0\le\rho\le 1$, is close to  
the standard Kuramoto order parameter $Z$ defined as
\[
Z=N^{-1}\sum_{k=1}^N e^{\ii\vp_k}.
\]
Indeed, substituting here the M\"obius transformation \eqref{eq:mtr}, 
we obtain~\cite{Pikovsky-Rosenblum-08}
\be
\label{eq:zZrel}
\begin{aligned}
Z&=N^{-1}\sum_{k=1}^N \frac{z+e^{\ii \psi_k}}{1+z^*e^{\ii \psi_k}}\\
&=z-\sum_{m=2}^\infty (1-|z|^2)(-1)^m (z^*)^{m-1}C_m\;,
\end{aligned}
\ee 
where $C_m=N^{-1}\sum_{k} \exp[\ii m\psi_k]$ are amplitudes of Fourier modes of 
the distribution of the WS phases $\psi_k$. 
Expression \eqref{eq:zZrel} is valid for $|z|<1$.
One can see that generally $z$ deviates from $Z$, although
their values coincide in the asynchronous state $z=Z=0$.
In the fully synchronous state 
one should use directly Eq.~\eqref{eq:mtr} which yields that for $|z|=1$ either 
all phases $\vp_k$ coincide, i.e. also $|Z|=1$, or at most one oscillator deviates from
the fully synchronous cluster. A special case corresponds to the uniform distribution
of the WS phases $\psi_k$, i.e. to the situation where all $C_m$ with $m\neq kN$
vanish ($k\ne 0$ is an arbitrary integer).
Then $z= Z$ in the thermodynamic limit $N\to\infty$, while for a finite $N$ there
are corrections $\sim |z|^N$. This allows one to consider the WS equation \eqref{eq:newz}
as  the one for the Kuramoto order parameter $Z$. For the Kuramoto-Sakaguchi problem, where
$H=e^{i\beta}Z$, this yields the closed equation for the order
parameter, first derived in a different way by 
Ott and Antonsen~\cite{Ott-Antonsen-08}. 
An essential part of our perturbation approach below deals with the distribution
of the WS phases, in fact we show that due to inhomogeneities and noise
it is close to the uniform one.
A final remark on the meaning of the new variables: the third WS variable, 
angle $\alpha$, determines shift of individual oscillators with respect 
to $\mbox{arg}(Z)$ and is not important for the collective dynamics.

Analysis of Eqs.~(\ref{eq:newz},\ref{eq:phased},\ref{eq:b}) is not an 
easy task, even under assumption that the perturbation terms $F_k$ are small. 
The main difficulty is that generally the WS complex amplitude  
$z=\rho\exp[\ii\Phi]$ is time-dependent. 
Therefore below we restrict ourselves to the case where in the absence 
of perturbations $\rho=\mbox{const}$ and $\dot\Phi=\mbox{const}$. 
Such a regime with $\rho\neq 1$ appears at least in two situations 
that attracted 
large interest recently. 
The first one is the chimera state 
(see Refs.~\cite{Kuramoto-Battogtokh-02,Abrams-Mirollo-Strogatz-Wiley-08}),
where a part of the population is fully synchronous, while the other part
is not. The latter sub-population is quantified by a uniformly rotating 
WS complex amplitude $z$, what implies $\rho=\mbox{const}$, $0<\rho<1$
\footnote{Generally, more complicated, quasiperiodic chimera states are 
also possible.}. 
Another
situation is partial synchrony due to nonlinear coupling, 
described in Refs.~\cite{Rosenblum-Pikovsky-07,*Pikovsky-Rosenblum-09}. 
In both cases, there exists a
rotating reference frame where the WS complex amplitude is constant, i.e.
$\rho\exp[\ii\Phi]=\mbox{const}$. In this frame, for the unperturbed state
the quantities $\w,H$ are constants and satisfy 
$H=-\ii 2\omega\rho(1+\rho^2)^{-1}e^{\ii\Phi}$. 
Below we consider perturbations to this state. 

It is convenient to introduce shifted WS phases according to 
$\psi_k=\theta_k+\Phi$ and to write $P= (Q+\ii S)e^{\ii\Phi}$,
then the system (\ref{eq:newz},\ref{eq:phased},\ref{eq:b})
can be re-written as a system of real equations
\be\label{eq:statsys}
\begin{aligned}
\dot{\theta_k}=&
\frac{F_k}{1-\rho^2}\left[2\rho\cos\theta_k+1+\rho^2\right]		
-\\&-
2\frac{S\rho+S\cos\theta_k-Q\sin\theta_k}{1-\rho^2}+\Omega\;,\\
S=&\frac{1}{1-X^2-Y^2}\frac{1}{N}\sum_{k=1}^N
F_k(2\rho\cos\theta_k+1+\rho^2)\\&\cdot 
(\cos\theta_k - X \cos\theta_k - Y \sin\theta_k)\;,\\
Q=&\frac{-1}{1-X^2-Y^2}\frac{1}{N}\sum_{k=1}^N
F_k(2\rho\cos\theta_k+1+\rho^2)\\&\cdot 
(\sin\theta_k-Y\cos\theta_k+X\sin\theta_k)\;,
\end{aligned}
\ee
where $\Omega=\omega\frac{1-\rho^2}{1+\rho^2}$ and 
$X+\ii Y=N^{-1}\sum_k \exp[\ii 2\theta_k]$. 
Formally, Eqs.~\eqref{eq:statsys} is
a system of phase oscillators $\theta_k$ driven by forces $F_k$ and subject
to mean fields $X,Y,S,Q$. In the unperturbed case $F_k=0$ it reduces to a 
system of uncoupled uniformly rotating phase oscillators. 

Below we analyze Eqs.~\eqref{eq:statsys} for two types of perturbations. 
In the first setup we consider
purely deterministic perturbations of the driving terms $\w,H$, namely we take 
$F_k=\e \left(u_k+\text{Im}\left[(f_k+\ii h_k) e^{\ii(\Phi-\varphi_k)}\right]\right)$. Here
$u_k$ determine spreading of natural frequencies $\w+\e u_k$, 
while the terms $f_k,h_k$ describe
variation of the forcing $H$  for individual units (cf. Eq.~\eqref{eq:posc}). 
Parameter $\e$ explicitly quantifies the level of inhomogeneity of the system; 
in the following treatment it is assumed to be small. 
Expressing $\exp[-\ii\vp_k]$ via
the WS complex amplitude $z$ and phases $\theta_k$ according to~\eqref{eq:mtr},
we obtain
\[
\begin{gathered}
F_k=\e \left [ u_k+ f_k\frac{2\rho
+(\rho^2-1)\sin\theta_k}{2\rho\cos\theta_k+1+\rho^2}+\right.\\\left.
+ h_k\frac{2\rho +(\rho^2+1)\cos\theta_k}{2\rho\cos\theta_k+1+\rho^2}\right ]\;,
\end{gathered}
\] 
which should be substituted in~\eqref{eq:statsys}.

We analyze the resulting system in the thermodynamic limit $N\to\infty$. In this
limit
the perturbation terms $u_k,f_k,h_k$ are described by their distribution density
$W(u,f,h)$ (without any restriction we can assume $\avr{u}=\avr{f}=\avr{h}=0$);
furthermore we seek for a solution with constant  mean fields $X,Y,S,Q$. 
Then the system~\eqref{eq:statsys} can be solved self-consistently: we find
the stationary distribution of $\theta_k$ for the given values of the
perturbations $w(\theta|u,f,h)$,
and then calculate the mean fields $X,Y,S,Q$ according to 
\be
\label{eq:meanX}
X=\iiint du\;df\;dh\; W(u,f,h)\int_0^{2\pi}d\theta\;w(\theta|u,f,h)\cos2\theta\;,
\ee
and similarly for other quantities. Since the expressions are lengthy, 
we present only the sketch of the derivation. 

One can see that the r.h.s. of the equation for $\dot\theta$ contains, together
with
constant terms, only terms $\sim\cos\theta,\;\sin\theta$, i.e.
$\dot\theta=A+B\cos\theta+C\sin\theta$. Thus, the stationary distribution of the
WS phases has the form $w(\theta|u,h,f)\sim
(A+B\cos\theta+C\sin\theta)^{-1}$. As a result, the integrals over $\theta$ in 
Eq.~\eqref{eq:meanX} (and in similar expressions for $Y,S,Q$) are reduced to 
solvable integrals of the type 
$\int_0^{2\pi}d\theta\cos (n\theta+\theta_0) (A+B\cos\theta+C\sin\theta)^{-1}$. 
This leads to rather lengthy but exact expressions, that however
can be expanded and simplified using the small parameter $\e$. 
The resulting formulas contain first and second powers of $u,f,h$, 
but the first powers disappear due to averaging with respect to density $W(u,f,h)$. 
The final formulas of the perturbation analysis, in the order $\sim\e^2$, are:
\be
\label{eq:pert}
\begin{aligned}
S=&
-\frac{\e^2}{2\W(1-\rho^2)}\left[2\rho(1+\rho^2)\avr{u^2}+\right.\\&+
\left.(4\rho^2+2\rho(\rho^2+1)\avr{u(f+h)}+
(1+\rho^2)^2\avr{uh}\right]\;,\\[1ex]
Q=& -\frac{\e^2}{2\W}\left[(1+\rho^2)\avr{uf}+2\rho\avr{f^2}+2\rho\avr{hf} 
 \right]\;,\\[1ex]
X=&\frac{\e^2}{
4\W^2(1-\rho^2)^2}[4\rho^2\avr{u^2}+
(\rho^2+1)^2\avr{h^2}+\\&+4\rho(\rho^2+1)\avr{hu}-(\rho^2-1)^2\avr{f^2}]\;,\\[1ex]
Y=&\e^2\frac{4\rho(\rho^2-1)\avr{uf}+(\rho^2+1)(\rho^2-1)\avr{hf}}{
4\W^2(1-\rho^2)^2}\;.
\end{aligned}
\ee

Let us consider two cases where these expressions simplify. If only the
natural frequency of oscillators $\w$ is distributed, but the force $H$ 
is the same for all oscillators,
then $u\neq 0$, $f=h=0$. In this case
\be
\label{eq:inhfr}
\begin{gathered}
S= -\frac{\e^2\rho(1+\rho^2)}{\W(1-\rho^2)}\avr{u^2},\qquad
X=\e^2\frac{\rho^2\avr{u^2}}{\W^2(1-\rho^2)^2},\\
Q=Y=0\;.
\end{gathered}
\ee
If the oscillators have the same frequency, but the sine-part of the force
varies, then
$u=h=0$ and $f\neq 0$, and we have
\be
\label{eq:inhforce}
Q=- \frac{\e^2\rho\avr{f^2}}{\W},\quad
X= -\frac{\e^2\avr{f^2}}{4\W^2},\quad
Y=S=0\;.
\ee
It is instructive to see how such perturbations look in terms of the 
original Eq.~(\ref{eq:newz}) for the WS complex amplitude $z$; 
this is accomplished by recalling that 
$P=(Q+\ii S)e^{\ii\Phi}$ and $z=\rho\exp[\ii\Phi]$, what yields
 \be
 \label{eq:pertz}
P= -\ii \e^2\frac{z(1+|z|^2)}{(1-|z|^2)\Omega}\langle  u^2\rangle,\quad
P=- \e^2\frac{z}{\Omega}\langle  f^2\rangle,
\ee
for the two considered cases. 

Finally, we express the Kuramoto order parameter
$Z=\avr{\exp[\ii\vp]}$ via the WS complex amplitude $z$,  using
general relation \eqref{eq:zZrel}.
For the inhomogeneous population we see that generally the amplitudes of 
the second harmonics $C_2=X+\ii Y$ of this distribution are non-zero, 
$X,Y\sim\e^2$  (higher harmonics have
higher orders in $\e$). Hence, the distribution of $\theta$ 
is non-uniform, though the corrections are small, $\sim\e^2$:
\[
Z= \rho
e^{\ii\Phi}[1-(1-\rho^2)(X+\ii Y)]\;,
\]
what yields for the two considered cases, in the leading order,
\be
\label{eq:zcor}
\begin{aligned}
Z= &z\left[1-\frac{\e^2|z|^2}{(1-|z|^2)\Omega^2}\langle 
u^2\rangle\right]\;,\\
Z= &z\left[1+\e^2\frac{1-|z|^2}{4\Omega^2}\langle  f^2\rangle\right]\;.
\end{aligned}
\ee

As a second application of our approach, we consider noisy
perturbations to the oscillators dynamics, taking $F_k=\e\xi_k(t)$, where $\xi_k(t)$ is
 a Gaussian white noise, $\langle \xi(t)\xi(t')\rangle=2\delta(t-t')$. 
As above, we consider perturbation
to the state with constant complex amplitude $z$, thus our starting point are 
Eqs.~\eqref{eq:statsys}. Furthermore, we take the thermodynamic limit
$N\to\infty$, allowing us to express the mean fields $X,Y,S,Q$ as averages over
the distribution of the WS phases. 
Here it is convenient not to average the equations for $S,Q$ directly, but to find
these mean fields from the solution of the Fokker-Planck equation for the
distribution of the phase $\theta$, 
which follows straightforwardly from the Langevin equation
\be
\label{eq:lang}
\begin{aligned}
\dot\theta=&\Omega-2\frac{\rho S}{1-\rho^2}
-\cos\theta\frac{S}{1-\rho^2}+\sin\theta
\frac{Q}{1-\rho^2}+\\&+
\e\xi(t)\left(\frac{1+\rho^2+2\rho\cos\theta}{1-\rho^2}\right).
\end{aligned}
\ee
Looking for a stationary solution of the Fokker-Planck equation, we use
smallness of 
$\e$ and represent the stationary distribution density as $w=w_0+\e^2 w_{2r}\cos
2\theta+\e^2w_{2i}\sin 2\theta$ (the expansion starts from the second harmonics terms 
because of the condition
$\avr{\exp[\ii\theta]}=0$).
This leads to the following expressions for $S,Q,w_{2r},w_{2i}$:
\be
\label{eq:pertnoise}
\begin{gathered}
S=w_{2r}=0,\\ Q=-\e^2\frac{2\rho(1+\rho^2)}{1-\rho^2},
\qquad w_{2i}= \e^2\frac{2\rho^2}{(1-\rho^2)^2\Omega}w_0\;.
\end{gathered}
\ee
Again, it is instructive to express the result of the perturbation 
analysis in terms of Eq.~(\ref{eq:newz}) and to write a relation 
between the Kuramoto order parameter and the 
WS complex amplitude:
\be
\label{eq:pn}
P= -\e^2\frac{2z(1+|z|^2)}{1-|z|^2}\;,\quad
Z=z\left[1+\e^2\frac{2\ii|z|^2}{(1-|z|^2)\Omega}\right].
\ee

Equations (\ref{eq:pertz},\ref{eq:zcor},\ref{eq:pn}) are the main result of the 
perturbation theory. They provide a closed description of the nonideal populations
of oscillators, where typically the driving field $H$ explicitly depends on the 
Kuramoto order parameter $Z$.

We now discuss the results of the perturbation analysis. 
We have considered in details two situations, that have been
previously treated in the framework of the standard WS analysis, i.e. for identical units.
In both analyzed cases, for a
non-identity of parameters
of the oscillators and for a noisy driving, we obtained that the WS phases, which
have an arbitrarily distribution  in the non-perturbed case, 
tend to a nearly uniform distribution with corrections $\sim\e^2$.
This results in the approximate relation between the Kuramoto order parameter
and the WS complex mean field, 
which differ by a small deviation $\sim\e^2$. This means that for weakly
perturbed situations, the WS equation
can be used for the evolution of the Kuramoto order parameter, with account
of above computed corrections $\sim\e^2$. 
As discussed above (see also~\cite{Pikovsky-Rosenblum-08}), the uniform
distribution of the WS phases is the 
case where the WS equations reduce to the Ott-Antonsen
equations~\cite{Ott-Antonsen-08}; sometimes this set of WS phases is called
Ott-Antonsen manifold. Our perturbation
analysis shows that small inhomogeneities ``drive'' the ensemble of oscillators
to an $\e^2$-vicinity of the Ott-Antonsen manifold, but not exactly to it.

A possible direction of future research is consideration of more 
generic perturbations.
Of particular interest is the case when the interaction between units is more 
complex than pure sine-coupling, e.g., when the second harmonic terms 
$\sim\exp[-\ii 2\vp]$ are present. 
Such setup can also be treated in the framework of the developed 
perturbation analysis;
the results will be reported elsewhere.

\acknowledgments
AP was supported by The Ministry of Education and
Science of the Russian Federation (the agreement of
August 27, 2013 N 02.B.49.21.0003 with the 
Lobachevsky State University of Nizhni Novgorod).
V. V. thanks the IRTG 1740/TRP 2011/50151-0, funded by the DFG /FAPESP.


\begin{thebibliography}{32}%
\makeatletter
\providecommand \@ifxundefined [1]{%
 \@ifx{#1\undefined}
}%
\providecommand \@ifnum [1]{%
 \ifnum #1\expandafter \@firstoftwo
 \else \expandafter \@secondoftwo
 \fi
}%
\providecommand \@ifx [1]{%
 \ifx #1\expandafter \@firstoftwo
 \else \expandafter \@secondoftwo
 \fi
}%
\providecommand \natexlab [1]{#1}%
\providecommand \enquote  [1]{``#1''}%
\providecommand \bibnamefont  [1]{#1}%
\providecommand \bibfnamefont [1]{#1}%
\providecommand \citenamefont [1]{#1}%
\providecommand \href@noop [0]{\@secondoftwo}%
\providecommand \href [0]{\begingroup \@sanitize@url \@href}%
\providecommand \@href[1]{\@@startlink{#1}\@@href}%
\providecommand \@@href[1]{\endgroup#1\@@endlink}%
\providecommand \@sanitize@url [0]{\catcode `\\12\catcode `\$12\catcode
  `\&12\catcode `\#12\catcode `\^12\catcode `\_12\catcode `\%12\relax}%
\providecommand \@@startlink[1]{}%
\providecommand \@@endlink[0]{}%
\providecommand \url  [0]{\begingroup\@sanitize@url \@url }%
\providecommand \@url [1]{\endgroup\@href {#1}{\urlprefix }}%
\providecommand \urlprefix  [0]{URL }%
\providecommand \Eprint [0]{\href }%
\providecommand \doibase [0]{http://dx.doi.org/}%
\providecommand \selectlanguage [0]{\@gobble}%
\providecommand \bibinfo  [0]{\@secondoftwo}%
\providecommand \bibfield  [0]{\@secondoftwo}%
\providecommand \translation [1]{[#1]}%
\providecommand \BibitemOpen [0]{}%
\providecommand \bibitemStop [0]{}%
\providecommand \bibitemNoStop [0]{.\EOS\space}%
\providecommand \EOS [0]{\spacefactor3000\relax}%
\providecommand \BibitemShut  [1]{\csname bibitem#1\endcsname}%
\let\auto@bib@innerbib\@empty
%</preamble>
\bibitem [{\citenamefont {Strogatz}(2000)}]{Strogatz-00}%
  \BibitemOpen
  \bibfield  {author} {\bibinfo {author} {\bibfnamefont {S.~H.}\ \bibnamefont
  {Strogatz}},\ }\href@noop {} {\bibfield  {journal} {\bibinfo  {journal}
  {Physica D}\ }\textbf {\bibinfo {volume} {143}},\ \bibinfo {pages} {1}
  (\bibinfo {year} {2000})}\BibitemShut {NoStop}%
\bibitem [{\citenamefont {Acebr{\'o}n}\ \emph {et~al.}(2005)\citenamefont
  {Acebr{\'o}n}, \citenamefont {Bonilla}, \citenamefont {Vicente},
  \citenamefont {Ritort},\ and\ \citenamefont {Spigler}}]{Acebron-etal-05}%
  \BibitemOpen
  \bibfield  {author} {\bibinfo {author} {\bibfnamefont {J.~A.}\ \bibnamefont
  {Acebr{\'o}n}}, \bibinfo {author} {\bibfnamefont {L.~L.}\ \bibnamefont
  {Bonilla}}, \bibinfo {author} {\bibfnamefont {C.~J.~P.}\ \bibnamefont
  {Vicente}}, \bibinfo {author} {\bibfnamefont {F.}~\bibnamefont {Ritort}}, \
  and\ \bibinfo {author} {\bibfnamefont {R.}~\bibnamefont {Spigler}},\
  }\href@noop {} {\bibfield  {journal} {\bibinfo  {journal} {Rev. Mod. Phys.}\
  }\textbf {\bibinfo {volume} {77}},\ \bibinfo {pages} {137} (\bibinfo {year}
  {2005})}\BibitemShut {NoStop}%
\bibitem [{\citenamefont {Pikovsky}\ and\ \citenamefont
  {Rosenblum}(2015)}]{Pikovsky-Rosenblum-15}%
  \BibitemOpen
  \bibfield  {author} {\bibinfo {author} {\bibfnamefont {A.}~\bibnamefont
  {Pikovsky}}\ and\ \bibinfo {author} {\bibfnamefont {M.}~\bibnamefont
  {Rosenblum}},\ }\href {\doibase http://dx.doi.org/10.1063/1.4922971}
  {\bibfield  {journal} {\bibinfo  {journal} {Chaos}\ }\textbf {\bibinfo
  {volume} {25}},\ \bibinfo {eid} {097616} (\bibinfo {year}
  {2015})}\BibitemShut {NoStop}%
\bibitem [{\citenamefont {Wiesenfeld}\ \emph {et~al.}(1996)\citenamefont
  {Wiesenfeld}, \citenamefont {Colet},\ and\ \citenamefont
  {Strogatz}}]{Wiesenfeld-Colet-Strogatz-96}%
  \BibitemOpen
  \bibfield  {author} {\bibinfo {author} {\bibfnamefont {K.}~\bibnamefont
  {Wiesenfeld}}, \bibinfo {author} {\bibfnamefont {P.}~\bibnamefont {Colet}}, \
  and\ \bibinfo {author} {\bibfnamefont {S.~H.}\ \bibnamefont {Strogatz}},\
  }\href@noop {} {\bibfield  {journal} {\bibinfo  {journal} {Phys. Rev. Lett.}\
  }\textbf {\bibinfo {volume} {76}},\ \bibinfo {pages} {404} (\bibinfo {year}
  {1996})}\BibitemShut {NoStop}%
\bibitem [{\citenamefont {Nixon}\ \emph {et~al.}(2013)\citenamefont {Nixon},
  \citenamefont {Ronen}, \citenamefont {Friesem},\ and\ \citenamefont
  {Davidson}}]{Nixon_etal-13}%
  \BibitemOpen
  \bibfield  {author} {\bibinfo {author} {\bibfnamefont {M.}~\bibnamefont
  {Nixon}}, \bibinfo {author} {\bibfnamefont {E.}~\bibnamefont {Ronen}},
  \bibinfo {author} {\bibfnamefont {A.~A.}\ \bibnamefont {Friesem}}, \ and\
  \bibinfo {author} {\bibfnamefont {N.}~\bibnamefont {Davidson}},\ }\href
  {\doibase 10.1103/PhysRevLett.110.184102} {\bibfield  {journal} {\bibinfo
  {journal} {Phys. Rev. Lett.}\ }\textbf {\bibinfo {volume} {110}},\ \bibinfo
  {pages} {184102} (\bibinfo {year} {2013})}\BibitemShut {NoStop}%
\bibitem [{\citenamefont {Martens}\ \emph {et~al.}(2013)\citenamefont
  {Martens}, \citenamefont {Thutupalli}, \citenamefont {Fourri{\`e}re},\ and\
  \citenamefont {Hallatschek}}]{Martens_etal-13}%
  \BibitemOpen
  \bibfield  {author} {\bibinfo {author} {\bibfnamefont {E.~A.}\ \bibnamefont
  {Martens}}, \bibinfo {author} {\bibfnamefont {S.}~\bibnamefont {Thutupalli}},
  \bibinfo {author} {\bibfnamefont {A.}~\bibnamefont {Fourri{\`e}re}}, \ and\
  \bibinfo {author} {\bibfnamefont {O.}~\bibnamefont {Hallatschek}},\
  }\href@noop {} {\bibfield  {journal} {\bibinfo  {journal} {Proc. Natl. Acad.
  Sci.}\ }\textbf {\bibinfo {volume} {110}},\ \bibinfo {pages} {10563}
  (\bibinfo {year} {2013})}\BibitemShut {NoStop}%
\bibitem [{\citenamefont {Temirbayev}\ \emph {et~al.}(2012)\citenamefont
  {Temirbayev}, \citenamefont {Zhanabaev}, \citenamefont {Tarasov},
  \citenamefont {Ponomarenko},\ and\ \citenamefont
  {Rosenblum}}]{Temirbayev_etal-12}%
  \BibitemOpen
  \bibfield  {author} {\bibinfo {author} {\bibfnamefont {A.~A.}\ \bibnamefont
  {Temirbayev}}, \bibinfo {author} {\bibfnamefont {Z.~Z.}\ \bibnamefont
  {Zhanabaev}}, \bibinfo {author} {\bibfnamefont {S.~B.}\ \bibnamefont
  {Tarasov}}, \bibinfo {author} {\bibfnamefont {V.~I.}\ \bibnamefont
  {Ponomarenko}}, \ and\ \bibinfo {author} {\bibfnamefont {M.}~\bibnamefont
  {Rosenblum}},\ }\href@noop {} {\bibfield  {journal} {\bibinfo  {journal}
  {Phys. Rev. E}\ }\textbf {\bibinfo {volume} {85}},\ \bibinfo {pages} {015204}
  (\bibinfo {year} {2012})}\BibitemShut {NoStop}%
\bibitem [{\citenamefont {Grollier}\ \emph {et~al.}(2006)\citenamefont
  {Grollier}, \citenamefont {Cros},\ and\ \citenamefont
  {Fert}}]{Grollier-Cros-Fert-06}%
  \BibitemOpen
  \bibfield  {author} {\bibinfo {author} {\bibfnamefont {J.}~\bibnamefont
  {Grollier}}, \bibinfo {author} {\bibfnamefont {V.}~\bibnamefont {Cros}}, \
  and\ \bibinfo {author} {\bibfnamefont {A.}~\bibnamefont {Fert}},\ }\href@noop
  {} {\bibfield  {journal} {\bibinfo  {journal} {Phys. Rev. B}\ }\textbf
  {\bibinfo {volume} {73}},\ \bibinfo {pages} {{060409(R)}} (\bibinfo {year}
  {2006})}\BibitemShut {NoStop}%
\bibitem [{\citenamefont {Heinrich}\ \emph {et~al.}(2011)\citenamefont
  {Heinrich}, \citenamefont {Ludwig}, \citenamefont {Qian}, \citenamefont
  {Kubala},\ and\ \citenamefont {Marquardt}}]{Heinrich_etal-11}%
  \BibitemOpen
  \bibfield  {author} {\bibinfo {author} {\bibfnamefont {G.}~\bibnamefont
  {Heinrich}}, \bibinfo {author} {\bibfnamefont {M.}~\bibnamefont {Ludwig}},
  \bibinfo {author} {\bibfnamefont {J.}~\bibnamefont {Qian}}, \bibinfo {author}
  {\bibfnamefont {B.}~\bibnamefont {Kubala}}, \ and\ \bibinfo {author}
  {\bibfnamefont {F.}~\bibnamefont {Marquardt}},\ }\href@noop {} {\bibfield
  {journal} {\bibinfo  {journal} {Phys. Rev. Lett.}\ }\textbf {\bibinfo
  {volume} {107}},\ \bibinfo {pages} {043603} (\bibinfo {year}
  {2011})}\BibitemShut {NoStop}%
\bibitem [{\citenamefont {Kiss}\ \emph {et~al.}(2002)\citenamefont {Kiss},
  \citenamefont {Zhai},\ and\ \citenamefont {Hudson}}]{Kiss-Zhai-Hudson-02a}%
  \BibitemOpen
  \bibfield  {author} {\bibinfo {author} {\bibfnamefont {I.}~\bibnamefont
  {Kiss}}, \bibinfo {author} {\bibfnamefont {Y.}~\bibnamefont {Zhai}}, \ and\
  \bibinfo {author} {\bibfnamefont {J.}~\bibnamefont {Hudson}},\ }\href@noop {}
  {\bibfield  {journal} {\bibinfo  {journal} {Science}\ }\textbf {\bibinfo
  {volume} {296}},\ \bibinfo {pages} {1676} (\bibinfo {year}
  {2002})}\BibitemShut {NoStop}%
\bibitem [{\citenamefont {Richard}\ \emph {et~al.}(1996)\citenamefont
  {Richard}, \citenamefont {Bakker}, \citenamefont {Teusink}, \citenamefont
  {Dam},\ and\ \citenamefont
  {Westerhoff}}]{Richard-Bakker-Teusink-Van-Dam-Westerhoff-96}%
  \BibitemOpen
  \bibfield  {author} {\bibinfo {author} {\bibfnamefont {P.}~\bibnamefont
  {Richard}}, \bibinfo {author} {\bibfnamefont {B.~M.}\ \bibnamefont {Bakker}},
  \bibinfo {author} {\bibfnamefont {B.}~\bibnamefont {Teusink}}, \bibinfo
  {author} {\bibfnamefont {K.~V.}\ \bibnamefont {Dam}}, \ and\ \bibinfo
  {author} {\bibfnamefont {H.~V.}\ \bibnamefont {Westerhoff}},\ }\href@noop {}
  {\bibfield  {journal} {\bibinfo  {journal} {Eur. J. Biochem.}\ }\textbf
  {\bibinfo {volume} {235}},\ \bibinfo {pages} {238} (\bibinfo {year}
  {1996})}\BibitemShut {NoStop}%
\bibitem [{\citenamefont {Breakspear}\ \emph {et~al.}(2010)\citenamefont
  {Breakspear}, \citenamefont {Heitmann},\ and\ \citenamefont
  {Daffertshofer}}]{Breakspear-Heitmann-Daffertshofer-10}%
  \BibitemOpen
  \bibfield  {author} {\bibinfo {author} {\bibfnamefont {M.}~\bibnamefont
  {Breakspear}}, \bibinfo {author} {\bibfnamefont {S.}~\bibnamefont
  {Heitmann}}, \ and\ \bibinfo {author} {\bibfnamefont {A.}~\bibnamefont
  {Daffertshofer}},\ }\href@noop {} {\bibfield  {journal} {\bibinfo  {journal}
  {{Frontiers in human neuroscience}}\ }\textbf {\bibinfo {volume} {{4}}},\
  \bibinfo {pages} {{190}} (\bibinfo {year} {{2010}})}\BibitemShut {NoStop}%
\bibitem [{\citenamefont {Prindle}\ \emph {et~al.}(2012)\citenamefont
  {Prindle}, \citenamefont {Samayoa}, \citenamefont {Razinkov}, \citenamefont
  {Danino}, \citenamefont {Tsimring},\ and\ \citenamefont
  {Hasty}}]{Prindle_etal-12}%
  \BibitemOpen
  \bibfield  {author} {\bibinfo {author} {\bibfnamefont {A.}~\bibnamefont
  {Prindle}}, \bibinfo {author} {\bibfnamefont {P.}~\bibnamefont {Samayoa}},
  \bibinfo {author} {\bibfnamefont {I.}~\bibnamefont {Razinkov}}, \bibinfo
  {author} {\bibfnamefont {T.}~\bibnamefont {Danino}}, \bibinfo {author}
  {\bibfnamefont {L.~S.}\ \bibnamefont {Tsimring}}, \ and\ \bibinfo {author}
  {\bibfnamefont {J.}~\bibnamefont {Hasty}},\ }\href@noop {} {\bibfield
  {journal} {\bibinfo  {journal} {Nature}\ }\textbf {\bibinfo {volume} {481}},\
  \bibinfo {pages} {39} (\bibinfo {year} {2012})}\BibitemShut {NoStop}%
\bibitem [{\citenamefont {N{\'e}da}\ \emph {et~al.}(2000)\citenamefont
  {N{\'e}da}, \citenamefont {Ravasz}, \citenamefont {Brechet}, \citenamefont
  {Vicsek},\ and\ \citenamefont
  {Barab{\'a}si}}]{Neda-Ravasz-Brechet-Vicsek-Barabasi-00}%
  \BibitemOpen
  \bibfield  {author} {\bibinfo {author} {\bibfnamefont {Z.}~\bibnamefont
  {N{\'e}da}}, \bibinfo {author} {\bibfnamefont {E.}~\bibnamefont {Ravasz}},
  \bibinfo {author} {\bibfnamefont {Y.}~\bibnamefont {Brechet}}, \bibinfo
  {author} {\bibfnamefont {T.}~\bibnamefont {Vicsek}}, \ and\ \bibinfo {author}
  {\bibfnamefont {A.-L.}\ \bibnamefont {Barab{\'a}si}},\ }\href@noop {}
  {\bibfield  {journal} {\bibinfo  {journal} {Nature}\ }\textbf {\bibinfo
  {volume} {403}},\ \bibinfo {pages} {849} (\bibinfo {year}
  {2000})}\BibitemShut {NoStop}%
\bibitem [{\citenamefont {Strogatz}\ \emph {et~al.}(2005)\citenamefont
  {Strogatz}, \citenamefont {Abrams}, \citenamefont {McRobie}, \citenamefont
  {Eckhardt},\ and\ \citenamefont {Ott}}]{Strogatz_et_al-05}%
  \BibitemOpen
  \bibfield  {author} {\bibinfo {author} {\bibfnamefont {S.~H.}\ \bibnamefont
  {Strogatz}}, \bibinfo {author} {\bibfnamefont {D.~M.}\ \bibnamefont
  {Abrams}}, \bibinfo {author} {\bibfnamefont {A.}~\bibnamefont {McRobie}},
  \bibinfo {author} {\bibfnamefont {B.}~\bibnamefont {Eckhardt}}, \ and\
  \bibinfo {author} {\bibfnamefont {E.}~\bibnamefont {Ott}},\ }\href@noop {}
  {\bibfield  {journal} {\bibinfo  {journal} {Nature}\ }\textbf {\bibinfo
  {volume} {438}},\ \bibinfo {pages} {43} (\bibinfo {year} {2005})}\BibitemShut
  {NoStop}%
\bibitem [{\citenamefont {Kuramoto}(1975)}]{Kuramoto-75}%
  \BibitemOpen
  \bibfield  {author} {\bibinfo {author} {\bibfnamefont {Y.}~\bibnamefont
  {Kuramoto}},\ }in\ \href@noop {} {\emph {\bibinfo {booktitle} {International
  Symposium on Mathematical Problems in Theoretical Physics}}},\ \bibinfo
  {editor} {edited by\ \bibinfo {editor} {\bibfnamefont {H.}~\bibnamefont
  {Araki}}}\ (\bibinfo  {publisher} {Springer Lecture Notes Phys., v. 39},\
  \bibinfo {address} {New York},\ \bibinfo {year} {1975})\ p.\ \bibinfo {pages}
  {420}\BibitemShut {NoStop}%
\bibitem [{\citenamefont {Kuramoto}(1984)}]{Kuramoto-84}%
  \BibitemOpen
  \bibfield  {author} {\bibinfo {author} {\bibfnamefont {Y.}~\bibnamefont
  {Kuramoto}},\ }\href@noop {} {\emph {\bibinfo {title} {Chemical Oscillations,
  Waves and Turbulence}}}\ (\bibinfo  {publisher} {Springer},\ \bibinfo
  {address} {Berlin},\ \bibinfo {year} {1984})\BibitemShut {NoStop}%
\bibitem [{\citenamefont {Sakaguchi}\ and\ \citenamefont
  {Kuramoto}(1986)}]{Sakaguchi-Kuramoto-86}%
  \BibitemOpen
  \bibfield  {author} {\bibinfo {author} {\bibfnamefont {H.}~\bibnamefont
  {Sakaguchi}}\ and\ \bibinfo {author} {\bibfnamefont {Y.}~\bibnamefont
  {Kuramoto}},\ }\href@noop {} {\bibfield  {journal} {\bibinfo  {journal}
  {Prog. Theor. Phys.}\ }\textbf {\bibinfo {volume} {76}},\ \bibinfo {pages}
  {576} (\bibinfo {year} {1986})}\BibitemShut {NoStop}%
\bibitem [{\citenamefont {Gupta}\ \emph {et~al.}(2014)\citenamefont {Gupta},
  \citenamefont {Campa},\ and\ \citenamefont {Ruffo}}]{Gupta-Campa-Ruffo-14}%
  \BibitemOpen
  \bibfield  {author} {\bibinfo {author} {\bibfnamefont {S.}~\bibnamefont
  {Gupta}}, \bibinfo {author} {\bibfnamefont {A.}~\bibnamefont {Campa}}, \ and\
  \bibinfo {author} {\bibfnamefont {S.}~\bibnamefont {Ruffo}},\ }\href@noop {}
  {\bibfield  {journal} {\bibinfo  {journal} {J. Stat. Mech. - Theor. Exp.}\
  }\textbf {\bibinfo {volume} {8}},\ \bibinfo {pages} {R08001} (\bibinfo {year}
  {2014})}\BibitemShut {NoStop}%
\bibitem [{\citenamefont {Rosenblum}\ and\ \citenamefont
  {Pikovsky}(2007)}]{Rosenblum-Pikovsky-07}%
  \BibitemOpen
  \bibfield  {author} {\bibinfo {author} {\bibfnamefont {M.}~\bibnamefont
  {Rosenblum}}\ and\ \bibinfo {author} {\bibfnamefont {A.}~\bibnamefont
  {Pikovsky}},\ }\href@noop {} {\bibfield  {journal} {\bibinfo  {journal}
  {Phys. Rev. Lett.}\ }\textbf {\bibinfo {volume} {98}},\ \bibinfo {pages}
  {064101} (\bibinfo {year} {2007})}\BibitemShut {NoStop}%
\bibitem [{\citenamefont {Pikovsky}\ and\ \citenamefont
  {Rosenblum}(2009)}]{Pikovsky-Rosenblum-09}%
  \BibitemOpen
  \bibfield  {author} {\bibinfo {author} {\bibfnamefont {A.}~\bibnamefont
  {Pikovsky}}\ and\ \bibinfo {author} {\bibfnamefont {M.}~\bibnamefont
  {Rosenblum}},\ }\href@noop {} {\bibfield  {journal} {\bibinfo  {journal}
  {Physica D}\ }\textbf {\bibinfo {volume} {238(1)}},\ \bibinfo {pages} {27}
  (\bibinfo {year} {2009})}\BibitemShut {NoStop}%
\bibitem [{\citenamefont {Kuramoto}\ and\ \citenamefont
  {Battogtokh}(2002)}]{Kuramoto-Battogtokh-02}%
  \BibitemOpen
  \bibfield  {author} {\bibinfo {author} {\bibfnamefont {Y.}~\bibnamefont
  {Kuramoto}}\ and\ \bibinfo {author} {\bibfnamefont {D.}~\bibnamefont
  {Battogtokh}},\ }\href@noop {} {\bibfield  {journal} {\bibinfo  {journal}
  {Nonlinear Phenom. Complex Syst.}\ }\textbf {\bibinfo {volume} {5}},\
  \bibinfo {pages} {380} (\bibinfo {year} {2002})}\BibitemShut {NoStop}%
\bibitem [{\citenamefont {Abrams}\ \emph {et~al.}(2008)\citenamefont {Abrams},
  \citenamefont {Mirollo}, \citenamefont {Strogatz},\ and\ \citenamefont
  {Wiley}}]{Abrams-Mirollo-Strogatz-Wiley-08}%
  \BibitemOpen
  \bibfield  {author} {\bibinfo {author} {\bibfnamefont {D.~M.}\ \bibnamefont
  {Abrams}}, \bibinfo {author} {\bibfnamefont {R.}~\bibnamefont {Mirollo}},
  \bibinfo {author} {\bibfnamefont {S.~H.}\ \bibnamefont {Strogatz}}, \ and\
  \bibinfo {author} {\bibfnamefont {D.~A.}\ \bibnamefont {Wiley}},\ }\href@noop
  {} {\bibfield  {journal} {\bibinfo  {journal} {Phys. Rev. Lett.}\ }\textbf
  {\bibinfo {volume} {101}},\ \bibinfo {pages} {084103} (\bibinfo {year}
  {2008})}\BibitemShut {NoStop}%
\bibitem [{\citenamefont {Watanabe}\ and\ \citenamefont
  {Strogatz}(1993)}]{Watanabe-Strogatz-93}%
  \BibitemOpen
  \bibfield  {author} {\bibinfo {author} {\bibfnamefont {S.}~\bibnamefont
  {Watanabe}}\ and\ \bibinfo {author} {\bibfnamefont {S.~H.}\ \bibnamefont
  {Strogatz}},\ }\href@noop {} {\bibfield  {journal} {\bibinfo  {journal}
  {Phys. Rev. Lett.}\ }\textbf {\bibinfo {volume} {70}},\ \bibinfo {pages}
  {2391} (\bibinfo {year} {1993})}\BibitemShut {NoStop}%
\bibitem [{\citenamefont {Watanabe}\ and\ \citenamefont
  {Strogatz}(1994)}]{Watanabe-Strogatz-94}%
  \BibitemOpen
  \bibfield  {author} {\bibinfo {author} {\bibfnamefont {S.}~\bibnamefont
  {Watanabe}}\ and\ \bibinfo {author} {\bibfnamefont {S.~H.}\ \bibnamefont
  {Strogatz}},\ }\href@noop {} {\bibfield  {journal} {\bibinfo  {journal}
  {Physica D}\ }\textbf {\bibinfo {volume} {74}},\ \bibinfo {pages} {197}
  (\bibinfo {year} {1994})}\BibitemShut {NoStop}%
\bibitem [{\citenamefont {Pikovsky}\ and\ \citenamefont
  {Rosenblum}(2008)}]{Pikovsky-Rosenblum-08}%
  \BibitemOpen
  \bibfield  {author} {\bibinfo {author} {\bibfnamefont {A.}~\bibnamefont
  {Pikovsky}}\ and\ \bibinfo {author} {\bibfnamefont {M.}~\bibnamefont
  {Rosenblum}},\ }\href@noop {} {\bibfield  {journal} {\bibinfo  {journal}
  {Phys. Rev. Lett.}\ }\textbf {\bibinfo {volume} {101}},\ \bibinfo {pages}
  {264103} (\bibinfo {year} {2008})}\BibitemShut {NoStop}%
\bibitem [{\citenamefont {Marvel}\ \emph {et~al.}(2009)\citenamefont {Marvel},
  \citenamefont {Mirollo},\ and\ \citenamefont
  {Strogatz}}]{Marvel-Mirollo-Strogatz-09}%
  \BibitemOpen
  \bibfield  {author} {\bibinfo {author} {\bibfnamefont {S.~A.}\ \bibnamefont
  {Marvel}}, \bibinfo {author} {\bibfnamefont {R.~E.}\ \bibnamefont {Mirollo}},
  \ and\ \bibinfo {author} {\bibfnamefont {S.~H.}\ \bibnamefont {Strogatz}},\
  }\href@noop {} {\bibfield  {journal} {\bibinfo  {journal} {Chaos}\ }\textbf
  {\bibinfo {volume} {19}},\ \bibinfo {pages} {043104.} (\bibinfo {year}
  {2009})}\BibitemShut {NoStop}%
\bibitem [{\citenamefont {Pikovsky}\ and\ \citenamefont
  {Rosenblum}(2011)}]{Pikovsky-Rosenblum-11}%
  \BibitemOpen
  \bibfield  {author} {\bibinfo {author} {\bibfnamefont {A.}~\bibnamefont
  {Pikovsky}}\ and\ \bibinfo {author} {\bibfnamefont {M.}~\bibnamefont
  {Rosenblum}},\ }\href@noop {} {\bibfield  {journal} {\bibinfo  {journal}
  {Physica D}\ }\textbf {\bibinfo {volume} {240}},\ \bibinfo {pages} {872}
  (\bibinfo {year} {2011})}\BibitemShut {NoStop}%
\bibitem [{\citenamefont {Braun}\ \emph {et~al.}(2012)\citenamefont {Braun},
  \citenamefont {Pikovsky}, \citenamefont {Matias},\ and\ \citenamefont
  {Colet}}]{Braun-etal-12}%
  \BibitemOpen
  \bibfield  {author} {\bibinfo {author} {\bibfnamefont {W.}~\bibnamefont
  {Braun}}, \bibinfo {author} {\bibfnamefont {A.}~\bibnamefont {Pikovsky}},
  \bibinfo {author} {\bibfnamefont {M.~A.}\ \bibnamefont {Matias}}, \ and\
  \bibinfo {author} {\bibfnamefont {P.}~\bibnamefont {Colet}},\ }\href@noop {}
  {\bibfield  {journal} {\bibinfo  {journal} {EPL}\ }\textbf {\bibinfo {volume}
  {99}},\ \bibinfo {pages} {20006} (\bibinfo {year} {2012})}\BibitemShut
  {NoStop}%
\bibitem [{\citenamefont {Vlasov}\ \emph {et~al.}(2015)\citenamefont {Vlasov},
  \citenamefont {Pikovsky},\ and\ \citenamefont
  {Macau}}]{Vlasov-Pikovsky-Macau-15}%
  \BibitemOpen
  \bibfield  {author} {\bibinfo {author} {\bibfnamefont {V.}~\bibnamefont
  {Vlasov}}, \bibinfo {author} {\bibfnamefont {A.}~\bibnamefont {Pikovsky}}, \
  and\ \bibinfo {author} {\bibfnamefont {E.~E.~N.}\ \bibnamefont {Macau}},\
  }\href {\doibase http://dx.doi.org/10.1063/1.4938400} {\bibfield  {journal}
  {\bibinfo  {journal} {Chaos}\ }\textbf {\bibinfo {volume} {25}},\ \bibinfo
  {eid} {123120} (\bibinfo {year} {2015})}\BibitemShut {NoStop}%
\bibitem [{\citenamefont {Ott}\ and\ \citenamefont
  {Antonsen}(2008)}]{Ott-Antonsen-08}%
  \BibitemOpen
  \bibfield  {author} {\bibinfo {author} {\bibfnamefont {E.}~\bibnamefont
  {Ott}}\ and\ \bibinfo {author} {\bibfnamefont {T.~M.}\ \bibnamefont
  {Antonsen}},\ }\href@noop {} {\bibfield  {journal} {\bibinfo  {journal}
  {CHAOS}\ }\textbf {\bibinfo {volume} {18}},\ \bibinfo {pages} {037113}
  (\bibinfo {year} {2008})}\BibitemShut {NoStop}%
\bibitem [{Note1()}]{Note1}%
  \BibitemOpen
  \bibinfo {note} {Generally, more complicated, quasiperiodic chimera states
  are also possible.}\BibitemShut {Stop}%
\end{thebibliography}
\end{document}